\documentclass[%
 aip,
 amsmath,amssymb,
 reprint,%
]{revtex4-1}

\usepackage{graphicx}% Include figure files
\usepackage{dcolumn}% Align table columns on decimal point
\usepackage{bm}% bold math
\usepackage{xcolor}

\usepackage[utf8]{inputenc}
\usepackage[T1]{fontenc}
\usepackage{mathptmx}

\begin{document}

\preprint{AIP/123-QED}

\title{Observation of mode splitting in artificial spin ice}
% Force line breaks with \\

\author{Sergi Lendinez}
\author{Mojtaba Taghipour Kaffash}%
 
\author{M. Benjamin Jungfleisch}
\affiliation{Department of Physics and Astronomy, University of Delaware, Newark, Delaware 19716, United States}
\email{mbj@udel.edu}

\begin{abstract}
We report the dependence of the magnetization dynamics in a square artificial spin-ice lattice on the in-plane magnetic field angle. Using two complementary measurement techniques -- broadband ferromagnetic resonance and micro-focused Brillouin light scattering spectroscopy -- we systematically study the evolution of the lattice dynamics, both for a coherent radiofrequency excitation and an incoherent thermal excitation of spin dynamics. We observe a splitting of modes facilitated by inter-element interactions that can be controlled by the external field angle and magnitude. Detailed time-dependent micromagnetic simulations reveal that the split modes are localized in different regions of the square network. This observation suggests that it is possible to disentangle modes with different spatial profiles by tuning the external field configuration. 
\end{abstract}

\maketitle

%\section{\label{sec:intro}Introduction}
Propagating spin waves (or their elementary quanta -- magnons) with wavelengths at the sub-micrometer length scale can carry and transport spin information in magnetic materials with low losses \cite{Csaba_2017}. Therefore, they have been discussed as data carriers in next-generation information technologies. Essential in this regard is magnonic crystals, which are artificially-designed periodic lattices in which the spin-wave band structure is engineered for optimized spin-wave properties. Hence, they are promising for applications in data processing, information technologies, and microwave devices \cite{Chumak_JPD_2017,Krawczyk_2014, Kruglyak_2010,Lenk_2011,Neusser_2009}. 
In recent years, artificial spin ices (ASIs) \cite{Wang2006,Farhan_2019,Drisko_2017} have been proposed as potential magnonic crystals, as they allow unprecedented reconfigurability, precise control of their ground state, and tuning of the magnetization dynamics \cite{Gliga_2020,Lendinez_2019,Skjaervo_2019}.

Micromagnetic modeling \cite{Gliga2013,Montoncello_2018,Stenning_2020,Mamica_2018,Iacocca2016,Ezio_2020,Dion_2019,Arroo_2019}, microwave spectroscopy \cite{Jungfleisch_PRB_Ice_2016,jungfleisch2017_PRA,Zhou2016,Bhat2016,Bhat2017,Bhat_PRL_2020,Talapatra_PRAppl_2020,Bang_JAP2019,Bang_PRB2019}, and Brillouin light scattering (BLS) spectroscopy \cite{Li_JAP2017,Mamica_2018,Bhat_PRL_2020,kaffash2020} have been used to gain a better understanding of the complex dynamics in ASI. Gliga et al. first proposed square ASI as a potential platform for spin-wave conduits \cite{Gliga2013}. First experimental efforts followed shortly after that and mainly relied on ferromagnetic resonance (FMR) spectroscopy \cite{Sklenar2013, Zhou2016, Bhat2016, Jungfleisch_PRB_Ice_2016}. Due to their high sensitivity, microwave-based techniques are well suited for detecting small signal levels such as those found in ASI. There has been an increased interest in the effects of dynamic mode hybridization in magnetic materials and nanostructures \cite{MacNeill_2019,Shiota_2020,Sud_2020,Li_PRL_2020} and, recently, high-resolution anticrossing spectra have been reported even in interacting square artificial spin ice \cite{Gartside_2021}. However, the high sensitivity comes at the cost of losing information about the spatial extent of the dynamics. 

Furthermore, by relying on a microwave antenna of finite size, one is limited to probing only a limited wavevector range. Additionally, microwave antennas couple efficiently only to odd spin-wave modes. Hence, they cannot be used to excite/detect even spin-wave modes reliably. As an alternative, micro-focused BLS has been used more recently to investigate the spatial distribution of microwave-driven spin dynamics in ASI \cite{kaffash2020,Bhat_PRL_2020}. On the other hand, wavevector-resolved BLS revealed the spin-wave dispersion of thermally excited ASI \cite{Li_JAP2017,Mamica_2018}. However, how the thermal spin-wave spectrum without any resonant microwave excitation compares to the traditionally-used microwave-spectroscopy techniques such as FMR, and even more importantly, how the dynamics and inter-element coupling depend on the in-plane magnetic field angle, have been open questions until now.

\begin{figure}[b]
\centering
\includegraphics[width=0.85\linewidth]{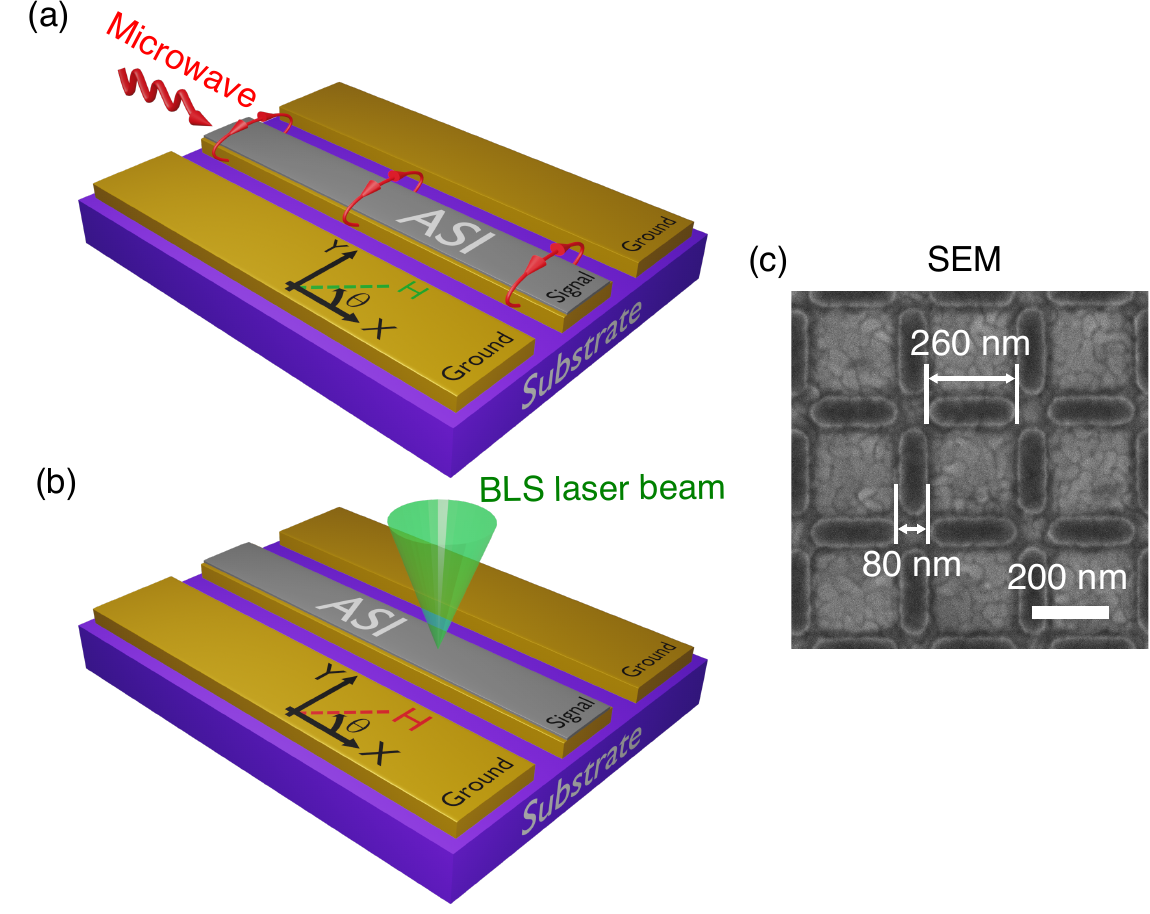}
\caption{\label{fig:setup} Illustration of the experimental setup and angular-dependent measurement configuration for (a) ferromagnetic resonance and (b) Brillouin light scattering measurement. The square artificial spin ice is patterned directly on the signal line (gray area). The bias magnetic field $\mu_0 H$ is applied in-plane at an angle $\theta$, where $\theta=0^\circ$ means along the $x$-axis, while $\theta=90^\circ$ corresponds to a field along the $y$-axis. (c) Scanning electron micrograph of the studied artificial spin ice with the dimensions of the nanomagnets.}
\end{figure}

Here, we study the detailed angular-dependent dynamic properties of a square artificial spin-ice lattice by three complementary techniques: we present a comparison of angular-dependent inductive FMR measurements with micro-focused BLS characterizations of thermally-excited spin dynamics. We observe a spin-wave mode splitting facilitated by the inter-element interactions that depends on both the in-plane field angle and magnitude. %We observe a splitting of the dynamic modes in the artificial spin ice that depends on both the in-plane field angle and magnitude. 
The experimentally-acquired spectra are interpreted using time-dependent micromagnetic simulations. Furthermore, the two-dimensional micromagnetic modeling results reveal that the split modes observed in the experiments reside in different regions of a single ASI vertex, suggesting that it is possible to control which portion of the network oscillates. %Our results show that ASIs are an ideal platform for magnonic devices.

Nanomagnetic islands with a size of 260 nm $\times$ 80 nm arranged on a square lattice with a lattice constant of 341 nm and a minimum gap of 34 nm have been fabricated. We patterned the lattice with electron beam lithography, deposited 20 nm of Ni$_{81}$Fe$_{19}$ using electron beam evaporation, followed by lift-off. The ASI was grown directly on top of the signal line of a coplanar waveguide that had previously been fabricated using optical lithography and electron beam evaporation of 150 nm of Au, see Figs.~\ref{fig:setup}(a,b). The signal line width is 20 $\mu$m, and the gap between the signal and ground lines is 10 $\mu$m. This geometry results in an ASI lattice composed of hundreds of thousands of elements, which we expect to behave as an infinitely-extended lattice. A scanning electron microscopy image of a section of the lattice with its dimensions is shown in Fig.~\ref{fig:setup}(c).

Two separate sets of measurements have been carried out: in-plane magnetic field dependent FMR and thermally-excited micro-focused BLS. For the FMR measurements, we use a vector network analyzer FMR (VNA-FMR) approach, in which a microwave signal is sent through the coplanar wave\-guide, and the transmitted signal is detected by the VNA (Keysight N5225A) measuring the transmission parameter S$_{12}$. A higher absorption at a given field and frequency indicates that the system magnetization is on resonance. Since there is an oscillating driving, the measurement is more sensitive when the magnetization is perpendicular to the oscillation direction (parallel to the signal line). At a given in-plane field angle, we sweep the magnetic field from negative to positive values, and we record the frequency-dependent S$_{12}$ parameter at each field step. To vary the in-plane field angle, the direction of the magnetic field is rotated with respect to the signal line.

\begin{figure}[t]
\centering
\includegraphics[width=0.65\linewidth]{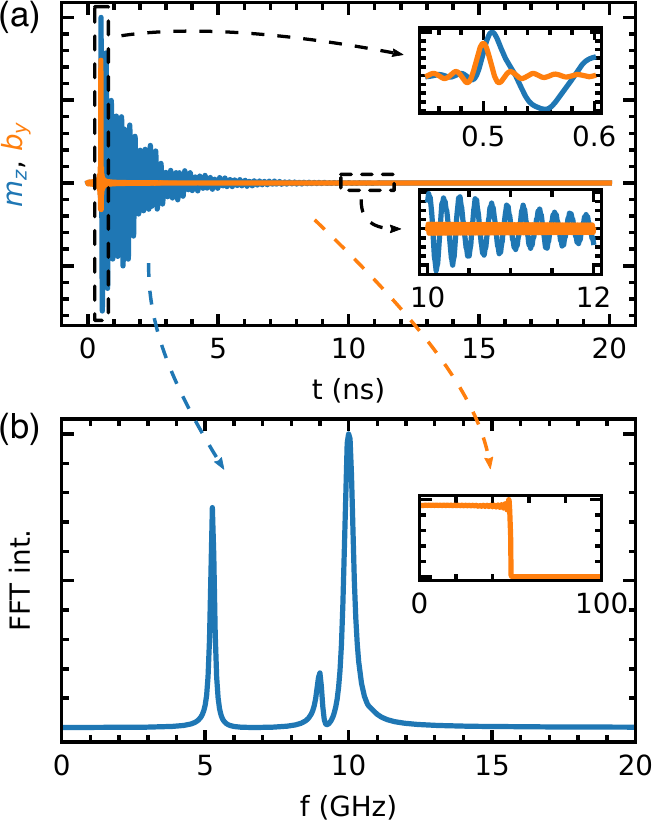}
\caption{\label{fig:sim} Illustration of the dynamic micromagnetic simulation approach. (a) Temporal evolution of the exciting magnetic field sinc-pulse $b_\mathrm{y}$ (orange) and the resulting onset of magnetization dynamics, shown here: $m_\mathrm{z}$ (blue). The insets show $m_\mathrm{z}$ and $b_\mathrm{y}$ on a magnified scale in the instance of time when  $b_\mathrm{y}$ is applied and long after that when a steady-state oscillation of $m_\mathrm{z}$ is observed.  (b) Corresponding Fourier transform shows the eigenexcitations of the square lattice at an external magnetic field of $B=2$ mT applied at an angle of $\theta=0^\circ$. The inset shows the Fourier transform of the sinc pulse shown in (a) that features a precise cut-off frequency of 50 GHz.}
\end{figure}

The inductive FMR measurements are compared to micro-focused-BLS studies, which reveal the thermally excited spectra at various in-plane field angles. For this purpose, a continuous single-mode 532-nm wavelength laser (Spectra Physics Excelsior) is focused on the surface of the same ASI lattice as studied by FMR. A sketch of the measurement configuration is shown in Fig.~\ref{fig:setup}(b). The BLS process can be understood by the inelastic scattering of laser photons with magnons. The BLS process is energy and momentum conserving, and thus the scattered photons carry information about the probed magnon \cite{Madami_2012}. In our micro-focused system, a high-numerical aperture (NA = 0.75) objective lens is used. Therefore, the scattered and reflected light is collimated within a large cone angle with respect to the sample surface normal resulting in the detection of an extensive wavevector range between 0 -- 17.8~rad/$\mu$m \cite{Sebastian_2015}. %ea range of momentums proportional to the aperture of the objective are simultaneously detected. 
In our setup, we can rotate the sample with respect to the external magnetic field to perform angular-dependent measurements. Since the magnetization is thermally excited, all magnetization directions are equally sensitive regardless of their orientation with respect to the signal line, which is in stark contrast to the typically-employed FMR technique. Additionally, unlike FMR, which relies on a coupling of the microwave drive with spin dynamics, both odd and even spin-wave modes can be probed with equal sensitivity in thermal-BLS measurements.

The experimental data are compared to micromagnetic simulations performed using MuMax3 \cite{mumax3}. The ASI is simulated using a single vertex and periodic boundary conditions, with the same geometry and dimensions as the nanofabricated sample. The simulation space is divided in $230 \times 230 \times 1$ cells with dimensions of $2.9 \times 2.9 \times 15$ nm$^3$. The lateral cell size is chosen to be smaller than 5 nm, corresponding to the exchange length of Ni$_{81}$Fe$_{19}$. Starting from an equilibrium state, the magnetization dynamics are simulated by sending an excitation sinc field pulse in the $y$-axis, with an amplitude of $b_y=0.1$ mT and cut-off frequency of 50 GHz [orange curve in Fig.~\ref{fig:sim}(a)]. The spatial distribution and the average magnetization components are saved for a total time of 20 ns during the excitation, as shown in blue for the averaged $z$-component in Fig.~\ref{fig:sim}(a) for an external field of $B=2$ mT. As can be seen in Fig.~\ref{fig:sim}(a), the magnetization starts precessing after the sinc pulse is applied (detailed in the upper inset). After some transition time, steady oscillations can be observed (lower inset). To gain further insight into the magnetic dynamics and determine the characteristic resonant frequencies, we perform a Fast Fourier Transform (FFT) to the time trace of the $z$-component of the magnetization. The FFT intensity is computed as the square power of the FFT amplitude, and the peaks in intensity correspond to the dominant oscillation frequencies, as shown in Fig.~\ref{fig:sim}(b) for the averaged $z$-component. The sinc pulse is chosen since its transformed spectrum is flat up to the cut-off frequency, at which point its amplitude decreases to almost 0, as is evidenced in the inset of Fig.~\ref{fig:sim}(b); this reduces a spurious contribution of characteristic frequencies in the FFT intensity spectrum associated with the pulse. Spatial profiles can be obtained similarly by performing the FFT of the time series of each cell. The FFT intensity spectra are hence obtained by simulating the time evolution in the field range of $-300$ mT to 300 mT.

\begin{figure}[t]
\centering
\includegraphics[width=1\linewidth]{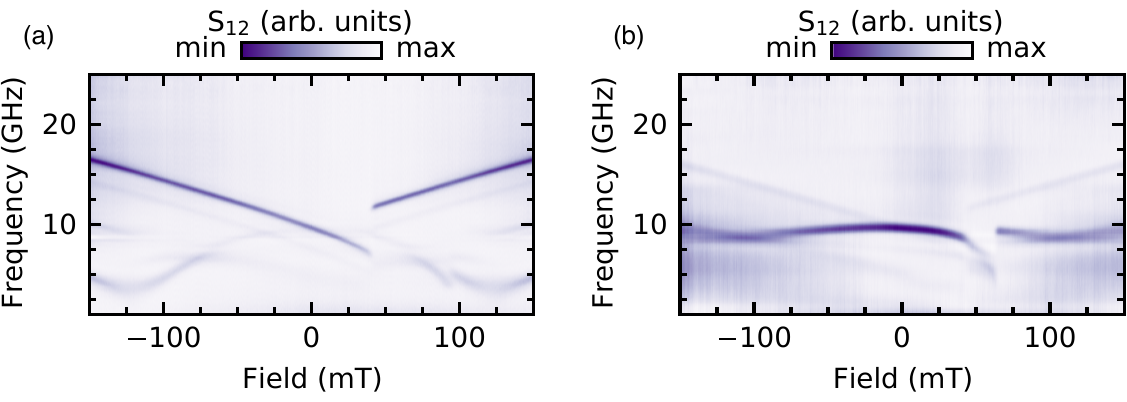}
\caption{Experimentally observed broadband ferromagnetic resonance spectra at magnetic fields applied at (a) $\theta=0^\circ$ and (b) $\theta=75^\circ$. The color-coded maps show the strongest microwave absorption corresponding to a minimum scattering parameter $S_\mathrm{12}$ in dark colors.}
\label{fig:fmr}
\end{figure}

In the following, we discuss the experimental results obtained by broadband FMR. Figure~\ref{fig:fmr} shows the experimental FMR data at applied magnetic field angles of $\theta=0^\circ$ and $\theta=75^\circ$. 
%Figure~\ref{fig:0deg} shows the spectra at external magnetic fields applied along the signal line, $\theta=0^\circ$. Experimental FMR data is presented in Fig.~\ref{fig:0deg}(a) and the corresponding results of thermally-excited micro-BLS is shown in Fig.~\ref{fig:0deg}(b). For comparison, we show micromagnetic simulations in Fig.~\ref{fig:0deg}(c,d) (in linear color scale and logarithmic color scale, respectively. 
The dark lines in the FMR spectra (Fig.~\ref{fig:fmr}) show a decrease in the transmission parameter S$_{12}$ indicating a resonant excitation of spin dynamics in the ASI. For an applied-field angle of $\theta=0^\circ$ [Fig.~\ref{fig:fmr}(a)], the most intense mode is characterized by a monotonous increase in frequency as the magnitude of the external field increases and is produced by the islands aligned along the signal line (in the following we refer to those islands as \textit{horizontal} islands). Fainter absorption lines running parallel to the main absorption lines but at a $\sim$2 GHz lower frequency are identified as edge modes residing in the horizontal islands \cite{Jungfleisch_PRB_Ice_2016}. At even lower frequencies, we observe a characteristic W-shaped line, with minima around $-175$ mT and $175$ mT, and a maximum of $\approx$ 10~GHz at 0 mT. This absorption line originates from the vertical islands (aligned perpendicular to the signal line), and, hence, it becomes weaker as the magnetic field approaches 0 mT and the magnetization is perpendicular to the signal line. In this situation, the microwave field is parallel to the magnetization leading to a reduced torque.

In order to tune the inter-element interactions, we change the magnetic field angle. Figure~\ref{fig:fmr}(b) shows the FMR spectra at applied magnetic fields of $\theta=75^\circ$. As is apparent from the figure, the most intense mode (predominantly associated with the dynamics in the horizontal islands) is lowered in frequency and is mostly flat; i.e., $\partial f/\partial H \approx 0$ in the investigated field range. This behavior is fundamentally different than what is observed at 0$^\circ$ [Fig.~\ref{fig:fmr}(a)], where the slope of the most intense curve $\partial f/\partial H$ monotonically decreases as the field is swept from negative to positive fields until the horizontal islands switch at around $+80$~mT. Moreover, we find indications that modes cross as we change the field angle. 

\begin{figure}[t]
\centering
\includegraphics[width=1\linewidth]{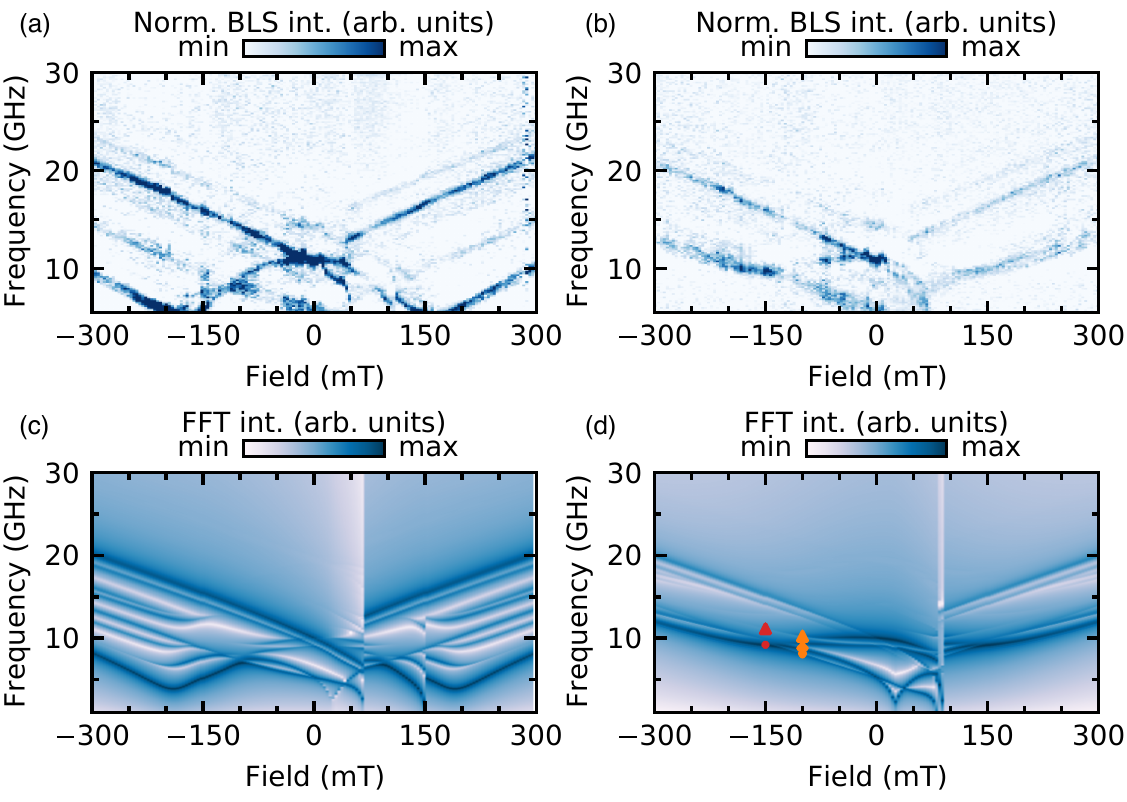}
\caption{\label{fig:BLS_sim} (a) Brillouin light scattering data for applied magnetic fields at $\theta=0^\circ$ and (b) $\theta=75^\circ$. In this case, dark color in the color-coded maps represents higher intensity. (c) Corresponding micromagnetic simulations at $\theta=0^\circ$ and (d) $\theta=75^\circ$.}
\end{figure}

To gain further insights, we employ a second experimental technique -- micro-focused BLS on thermally-excited spin waves.
Figures~\ref{fig:BLS_sim}(a,b) show the thermally-excited BLS spectra for 0$^\circ$ and 75$^\circ$, respectively. While the same general trend of the main modes is observed, we find a distinct difference in the BLS data. Significantly more modes are detectable in the thermal BLS spectra. Furthermore, the most prominent resonant lines -- coming from horizontal and vertical elements -- have comparable intensities. This can be understood from the fact that Figs.~\ref{fig:BLS_sim}(a,b) show the thermal BLS signal; i.e., the resonances are driven by thermal activation, which is independent of the relative orientation of the signal line (and thus the driving microwave magnetic field in FMR experiments) and the direction of the magnetization. Therefore, the torque on the vertical islands' net magnetization is not reduced as the field is approaching 0 mT. Moreover, additional modes, barely visible in the FMR data, can be detected by BLS. We will discuss this interesting observation below.

We now compare the experimental data with micromagnetic simulations, shown in Figs.~\ref{fig:BLS_sim}(c,d). As is evident from a comparison to the BLS data [Figs.~\ref{fig:BLS_sim}(a,b)], %an excellent agreement with the micromagnetic simulations is found. %The similarities between FMR and micromagnetics are best observed when the simulation data is plotted on a linear scale [Fig.~\ref{fig:0deg}(c)]. We detect many more modes by BLS than by FMR. To make the lower-intensity modes better visible, we plot the simulation results on logarithmic scale, Fig.~\ref{fig:0deg}(d). 
the agreement between the logarithmic color-coded micromagnetic simulation results and the micro-BLS data is remarkable. It highlights that micro-BLS is well suited for detecting most eigenmodes in the ASI network, even those inaccessible by standard broadband FMR.

%Since we can observe more modes in the micro-focused BLS data, we plot the simulated spectra using a logarithmic scale [Fig.~\ref{fig:0deg}(d)] in order to observe lower-intensity modes that are not visible in a linear scale. The agreement between micromagnetic simulations and both sets of data shows that our micromagnetic model captures most of the details of the lattice. In addition, the spatially-resolved micromagnetic simulations allow us to confirm the location of the excitation in the lattice.

\begin{figure}[t]
\centering
\includegraphics[width=0.75\linewidth]{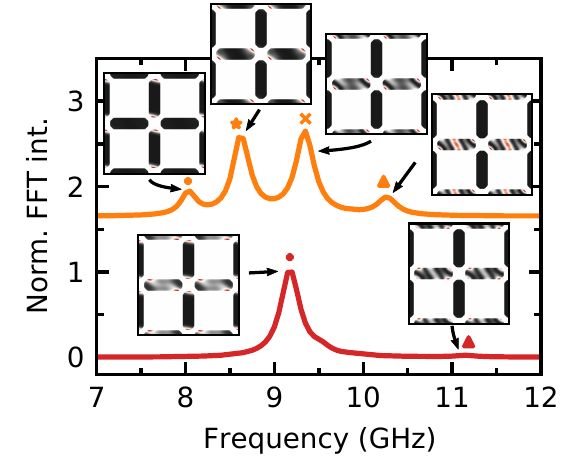}
\caption{\label{fig:profile} Simulated frequency response of the modes observed at $\theta=75^\circ$ for $-150$~mT (red, bottom curve) and $-100$~mT (orange, top curve). The spatial profiles of the observed modes are shown as insets as color coded images: red represents a high intensity and black a small intensity.}
\end{figure}

Inspecting the spectra obtained at an in-plane field angle of $\theta=0^\circ$ more closely, we can observe several modes that are crossing. For example, in the BLS [Fig.~\ref{fig:BLS_sim}(a)] and the corresponding simulated data [Fig.~\ref{fig:BLS_sim}(c)], we find such a mode crossing at a magnetic field of approximately $-100$ mT and a frequency of 8 GHz. This crossing is also visible in the FMR results [Fig.~\ref{fig:fmr}(a)]; however, the FMR signal of the intersecting modes is much weaker. %To better understand the previously demonstrated tunability of the inter-element interaction by varying the in-plane field angle in FMR and, thus, the crossing behavior, we perform the corresponding experiment by BLS and compare the results to micromagnetic modeling. 
 As discussed above, the FMR data shows a change in the lineshape and crossings as the in-plane field angle is changed. With the BLS technique, we are better equipped to understand the dynamics since we can detect more modes and are equally sensitive to both sublattices. 
By changing the direction of the external field to $\theta=75^\circ$ [Figs.~\ref{fig:BLS_sim}(b,d)], the field dependence of the main resonant line changes and, as a result, the crossing behavior is altered. At magnetic fields around $-100$ mT, we find that the main resonance splits in different modes. In particular, in the BLS data, we observe that the resonance splits into at least two modes with opposite slopes [Fig.~\ref{fig:BLS_sim}(b)], while in the micromagnetic simulated spectra, a mode splitting into four lines is found [Fig.~\ref{fig:BLS_sim}(d)]. The corresponding frequency response obtained in simulations is shown in Fig.~\ref{fig:profile} for two different field magnitudes of $-150$ mT (red curve) and $-100$ mT (orange curve) at 75$^\circ$: at a higher field magnitude, there is a high-intensity mode (red dot) and a very small-intensity mode (red triangle), while at the lower field magnitude the same modes appear to have split into four separated peaks (indicated by an orange dot, star, cross, and triangle). In the micromagnetic spectra, the mode splitting appears to be accompanied by a hybridization-like behavior indicated by avoided crossings \cite{Gartside_2021}. However, we cannot observe clear evidence of such a mode hybridization facilitated by inter-element coupling in the experiments -- neither in the BLS nor in FMR results. 
%From the sensitivity of our experimental FMR data we are not able to determine whether the inter-element coupling strength in our lattice sufficiently strong to facilitate a hybridization of modes \cite{Gartside_2021}. 

%In the data corresponding to applied fields of 0$^\circ$ we can observe mode intersections. For example, in the BLS and simulated data, we can observe this crossing at a magnetic field of $-100$ mT and frequency of 10 GHz. By changing the direction of the external field to $\theta=75^\circ$ (Fig.~\ref{fig:75deg}), the shape of the main resonant line changes and different crossings appear. At magnetic fields around $-100$ mT, we can observe the main resonance splitting. In the BLS data, the resonance splits into at least two modes with opposite slope. In the micromagnetic simulated spectra, it split into four lines. Figure 5 shows the spectrum at a field of $-150$ mT (blue curve), with only one main resonance, and at a field of $-100$ mT (orange curve), with four peaks in the intensity spectrum.

We conduct spatially-resolved micromagnetic simulations and determine the exact location of the excitations in the lattice, see insets in Fig.~\ref{fig:profile}. While not done here, we note that micro-focused BLS could potentially serve as an experimental method to record the spatially-resolved dynamics \cite{Bhat_PRL_2020,kaffash2020}. %A two-dimensional study of the excitation modes reveals the details of the mode splitting. 
Before the splitting occurs, at a magnetic field of $-150$ mT (red curve), the main excitation mode, centered at 9.2 GHz (marked with a red dot), is located on the horizontal islands and the edges of the vertical islands. Since the external field is applied at an angle of $\theta=75^\circ$, the mode in the horizontal islands is tilted, and the regions oscillating with a higher amplitude are pushed to the edge of the islands. A higher-order mode can also be observed at a higher frequency of 11.15 GHz, but with a much smaller intensity (marked with a red triangle). When the magnetic field is reduced to $-100$ mT (orange curve), we observe the splitting into four resonant modes. The lowest-frequency mode, centered at 8.05 GHz (marked with an orange dot), corresponds to the vertical islands' edge resonance mode, whereas the highest-frequency mode, centered at 10.25 GHz (marked with an orange triangle), corresponds to the higher-order excitation of the horizontal islands. Interestingly, the modes in between, at 8.6 GHz (marked with an orange star) and 9.35 GHz (marked with an orange cross), arise from the main mode located in the horizontal islands.

Hence, by varying the external field angle, we are able to split the main resonant mode into at least two separated resonant modes. The splitting comes with emerging oscillations located in different regions of the horizontal islands. Thus, we can control the frequency splitting of the resonances and their spatial resonant profiles. Moreover, since the slope of the frequency becomes almost flat, we show that it is in principle possible to change the oscillating regions by just varying the external magnetic field magnitude while keeping the excitation frequency nearly constant.

%Experimentally, we are able to observe this splitting in the FMR [Fig.~\ref{fig:fmr}(b)] and the BLS data [Fig.~\ref{fig:BLS_sim}(b)]. However, we are unable to experimentally determine the spatial location of these modes.

%Interestingly, we can observe avoided crossings in some of the simulated data, e.g. at $-175$ mT for applied magnetic fields at $\theta=75^\circ$. While this might be indications of hybridisation \cite{Gartside_2021}, we are unable to experimentally resolve these crossings.

In summary, we employed three different techniques -- broadband ferromagnetic resonance, micro-focused Brillouin light scattering, and micromagnetic simulations -- to investigate the angular-dependent spin dynamics in square artificial spin ice. We observe a splitting of modes that depends on both the magnetic field angle and magnitude. While the micromagnetic modeling results suggest that this splitting could be due to an avoided crossing or hybridization of modes, we cannot observe clear evidence of such a behavior in the experiments.
%This splitting could be interpreted as an avoided crossing or hybridization of modes. NOT SURE ABOUT THIS SENTENCE, SINCE WE THEN SAY IT'S NOT HYBRIDIZATION}. 
Using time-dependent micromagnetic modeling, we show that the mode splitting occurrence is due to oscillations localized in the different regions of a vertex and likely not an indication for mutual strong coupling of modes. This has direct consequences for interpreting the spectra and could potentially be utilized for a spatial disentanglement of modes.
In addition to the observed mode splitting, we demonstrate that micro-focused Brillouin light scattering is a powerful technique to explore the dynamic excitations in artificial spin ice. We show that the thermal spin-wave spectrum is sufficiently strong to be detected by micro-focused BLS and that using this approach, we can detect modes inaccessible by other commonly used methods such as microwave spectroscopy that rely on a resonant excitation. 

The data that support the findings of this study are available from the corresponding author upon reasonable request.

% study the detailed angular-dependent dynamic properties of a square artificial spin-ice lattice by three complementary techniques: we present a comparison of angular-dependent inductive FMR measurements with micro-focused BLS characterizations of thermally-excited spin dynamics. We observe a splitting of the dynamic modes in the artificial spin ice that depends on both the in-plane field angle and magnitude. The experimentally-acquired spectra are interpreted using time-dependent micromagnetic simulations. Furthermore, the two-dimensional micromagnetic modeling results reveal that the split modes observed in experiment reside in different regions of a single ASI vertex suggesting that it is possible to control which portion of the network oscillates while being nominally at the same resonance frequency. %Our results show that ASIs are an ideal platform for magnonic devices.
\begin{acknowledgments}
We thank Dr. Thomas Sebastian and Dr. Thomas Meyer for helpful discussions on BLS spectroscopy. We acknowledge technical help by Derrick Allen in setting up the laboratory. This work was supported by the U.S. Department of Energy, Office of Basic Energy Sciences, Division of Materials Sciences and Engineering under Award DE-SC0020308. %The authors also acknowledge support from the University of Delaware for collaboration with Argonne National Laboratory.
\end{acknowledgments}

%\nocite{*}
\bibliography{aipsamp}% Produces the bibliography via BibTeX.

\end{document}